\documentclass[%
onecolumn, prx, floatfix, longbibliography,superscriptaddress,
 amsmath,amssymb,
 aps,
]{revtex4-2}

\usepackage{graphicx}
\usepackage{dcolumn}
\usepackage{bm}
\usepackage{times}
\usepackage{hyperref}
\usepackage{physics}

\begin{document}
\preprint{}

\title {
Simple fusion-fission quantifies Israel-Palestine violence and suggests multi-adversary solution}

\author{Frank Yingjie Huo}
\affiliation{%
Dynamic Online Networks Laboratory, George Washington University, Washington, DC 20052, U.S.A.
}%
\author{Pedro D. Manrique}
\affiliation{%
Dynamic Online Networks Laboratory, George Washington University, Washington, DC 20052, U.S.A.
}%
\author{Dylan J. Restrepo}
\affiliation{%
International and Global Studies, Brandeis University, Waltham, MA 02453,  U.S.A.
}%
\author{Gordon Woo}
\affiliation{%
Moody's (Risk Management Solutions), London, EC3R 7BB U.K.
}%
\author{Neil F. Johnson$^*$}
\affiliation{%
Dynamic Online Networks Laboratory, George Washington University, Washington, DC 20052, U.S.A.
\\
$^*$  corresponding author: neiljohnson@gwu.edu}%



\baselineskip12pt


\begin{abstract}
{}
\end{abstract}

\maketitle

\onecolumngrid
\baselineskip24pt

{\bf 
Why humans fight has no easy answer. However, understanding better {\em how} humans fight could inform future interventions, hidden shifts and casualty risk. Fusion-fission describes the well-known grouping behavior of fish etc. fighting for survival in the face of strong opponents: they  form clusters (`fusion') which provide collective benefits and a cluster scatters when it senses danger (`fission'). Here we show how similar clustering (fusion-fission) of human fighters provides a unified quantitative explanation for complex casualty patterns across decades of Israel-Palestine region violence, as well as the October 7 surprise attack -- and uncovers a hidden post-October 7 shift. State-of-the-art data shows this fighter fusion-fission in action. It also predicts future `super-shock' attacks that will be more lethal than October 7 and will arrive earlier. It offers a multi-adversary solution. Our results -- which include testable formulae and a plug-and-play simulation -- enable concrete risk assessments of future casualties and policy-making grounded by fighter behavior.
}
\vskip0.2in

\section{Introduction}

Human conflict and terrorism pose a seemingly intractable challenge for governments, humanitarian organizations, development agencies and academia, and have attracted a wealth of valuable in-depth studies from myriad disciplinary perspectives  \cite{Israel1,Israel4,H4,H7,H8,Richardson,Mackay1,Mackay2,Mackay3,Gutfraind,Johnson,Johnson2,Peacock,Spagat2,Slaughter,Wrangham1,Wrangham2,Epstein,Kertesz,Gill,HH,Idriss,Overton,Horwood,sergey,Lazer,Axtell,Kalyvas,Cederman,singh,Strogatz,Kress,Science2011,Tivnan,Danforth,irwin,Strong,Storr,common,Hossack,Russell,Coulson,Robinson,Lucas,Hernandez,Armstrong,Berger}.
Part of this challenge is the urgent need of the trillion-dollar insurance industry and others, to better quantify the risk (i.e. probability) that a  future conflict/terrorism event will produce a large number of casualties. This translates mathematically into understanding the tails of a casualty distribution. The major complication is that a large casualty event like Hamas' 7 October, 2023 attack on Israel may never have happened before, or it is so rare that the data are sparse. A more mechanistic understanding of {\em how} such violent events are generated, could provide the insurance industry, policymakers and others with a concrete platform to explore what-if scenarios, potential impacts of interventions, and more rigorous  counterfactual thinking \cite{IUCRC,Gordon,GordonNeil,Gordon2}. Indeed, the need for innovation in this area is now so great that it has spawned the 2024 joint U.S.  National Science Foundation-government-industry initiative on terrorism and catastrophic online/cyber risks \cite{IUCRC}.

We are all familiar with footage showing how fish fight for survival in the face of strong opponents: they repeatedly cluster together 
(`fusion') then a cluster breaks up (`fission') when it senses danger \cite{Couzin,Levin}. Fusion brings collective benefits such as aggregated strength and awareness, while total fission (scattering) can be an effective response to imminent danger. Fusion-fission is ubiquitous across timescales, environments, geographical locations and species \cite{Couzin,Levin,Aureli}. 

Here we show how similar fusion-fission among human fighter forces can explain patterns in the violence across the Israel-Palestine region. These fighter forces include Hamas/Al-Qassam Brigades, Palestinian Islamic Jihad (PIJ), Fatah, Hezbollah, Al-Aqsa Martyrs' Brigades, Houthis, Islamic State (IS), Al-Nusrah Front -- all of which regard Israel as a strong opponent and all of which will typically need to adapt quickly in any fight to avoid annihilation \cite{Stijn}. This suggests a commonality of tactics that can then generate similar patterns in the  violence, as we find. Our mathematical analysis focuses on the mesoscale fighter cluster dynamics and does not need to specify why they fight or  individual-level identities, links or animosities, nor does the violence always need to directly involve Israel. We hence use the generic term `fighter force' instead of assigning labels like non-state army, terrorist organization, armed civilians, insurgency etc.; `overall fight' instead of war, small war, asymmetric war, conflict, civilian uprising, insurgency, terrorist campaign etc.; `fighter' instead of combatant, terrorist, armed civilian, non-state actor, insurgent, extremist, freedom fighter etc.; and `cluster' to denote some operationally cohesive unit of such fighters. As such, our findings can also help
deepen understanding of organizational behavior across categories of violence that can be hard to separate conceptually (e.g. terrorism vs. insurgency) \cite{common}. 

Our findings also suggest a counterintuitive policy takeaway: one of the world's most complex regions for violence and non-state armed actors (Israel-Palestine) may be among the simplest to understand in terms of how those fighters fight. Our quantitative results provide a rigorous foundation for  discussions of future interventions, hidden shifts and casualty risk -- and crucially, our mathematical approach and its results can be scrutinized and explored by any non-specialist via our plug-and-play fusion-fission simulation which requires no mathematical or coding knowledge:

\noindent \url{https://gwdonlab.github.io/netlogo-simulator/}

\begin{figure}[ht]
\includegraphics[width=0.9\linewidth]{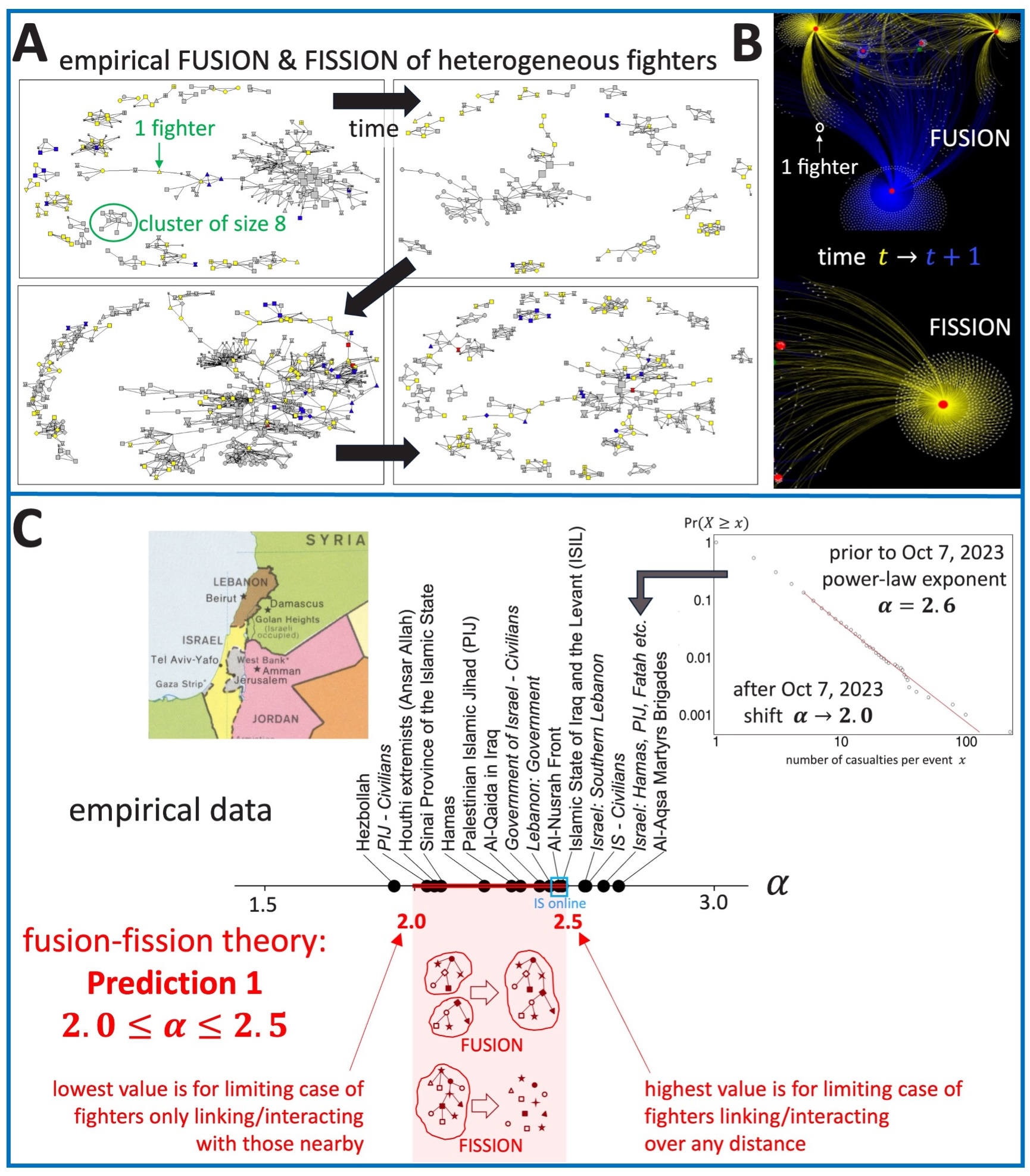}
    \caption{\bf Prediction 1 vs. empirical data. A: Consecutive time windows show clusters of fighters forming and breaking up (fusion-fission) in a fighter force that is a strong proxy for Israel's opponents \cite{B2B,Gill2}.  Fighter's (node) 
    attributes shown by shape (faction), size (role), color (skill). A link denotes two fighters (nodes) with a strong personal connection (see text) and hence interactions that promote operational cohesion  \cite{B2B,Gill2}.  B: Daily-scale fusion and fission among pro-IS (Islamic State) fighters online. Small white circle is 1 fighter (see SI Fig. 1 for typical profile). Colored circle: fighter community  \cite{PRL2023}. Links that disappear (appear) shown as yellow (blue).  C. Empirical exponent values $\alpha$ (black circles) for  approximate power-law casualty distributions (GED conflict data italics; GTD terrorism data non-italics) across Israel-Palestine region \cite{Stijn}. Prediction 1 matches this, i.e. $2.0\leq \alpha\leq 2.5$. Data-point for Gaza-West Bank casualties shifts to  $\alpha=2.0$ after 7 October 2023 (see text and Methods).}
    \label{fig:1}
\end{figure}

\section{Results}

\subsection{Fusion-fission empirical evidence}

Akin to footage of fish, the snapshots in
Fig. 1A show clusters of fighters (nodes) forming and breaking up. It uses data from a state-of-the-art study of fighter behavior (Provisional Irish Republican Army (PIRA) who fought rather successfully against a strong opponent (U.K.) who labelled them a terrorist organization) \cite{B2B,Gill2}. PIRA provide unique behavioral insight into fighter forces facing Israel  \cite{B2B,Gill2} because 
operational details often got copied  \cite{B2B,White,Frampton,
IRAPLO,groups}. Each fighter is a node and their faction (brigade), role and skill are denoted by the node's shape, size, and color respectively.
Two fighters are linked if they partnered in a prior operation and/or are close as friends, relatives or by marriage: each link hence facilitates operational interactions and cohesion
\cite{B2B,Gill2}. Reference \cite{B2B} confirms  these links/interactions gave PIRA strong operational cohesion. Each emerging cluster (e.g. size $s=8$ in  Fig. 1A) is a cohesive unit since its fighters are all directly or indirectly linked. A link says nothing about fighters' spatial proximity, since linked fighters can interact by phone or online and may create new links with distant fighters. 
Despite their army name and brigade structure, PIRA therefore exhibits cluster dynamics in time. Even though each snapshot aggregates over a finite time window, these cluster dynamics are strikingly similar to our simple plug-and-play fusion-fission simulation.

Figure 1B shows these cluster dynamics at daily-scale resolution. This time-lapse snapshot of pro-IS fighters operating in the online communications space  \cite{Science2016,PRL2023} reveals day-to-day fusion and total fission of clusters when in danger from security agencies. SI Fig. 1 illustrates how the individual fighter characteristics make them very different from casual anti-U.S. or anti-Israel etc. users \cite{Science2016}. Each link denotes a fighter (white node) being a member of a given online community (colored node) \cite{Science2016,PRL2023}. Studies show that  an online community's members feel strong links of trust with each other \cite{onlinetrust}. Hence each online community is a single cluster of linked fighters (i.e. cohesive unit) akin to those in Fig. 1A. Interestingly, the U.K.'s 2024 violent uprisings also featured  fighter (rioter) fusion-fission behavior online and offline
\cite{bbc1,bbc2}: their opponents (U.K. authorities) later confirmed that this fusion-fission made the fighters more unpredictable and hence harder to fight against \cite{bbc1,bbc2}.

\subsection{Fusion-fission mathematics and its predictions}
Our mathematics provides a mesoscale description of this cluster fusion-fission (Figs. 1A,B). The plug-and-play simulation shows it in real time. 
This mathematics is quite general and works by performing cluster-level averages: hence it is agnostic to the microscale details of which fighters and links are in what clusters; the precise nature of each operationally relevant link (e.g. trust, duty or something else); whether fighters leave the fighter force or become casualties and others join, as long as the fighter force size $N$ and its  heterogeneity  change slower than the fusion-fission rates \cite{PRL2023}; and whether the fusion and fission of clusters is spontaneous, pre-meditated, self-organized or managed, and its root causes. For example, for a given set of fusion-fission rates the fusion between large clusters may be pre-meditated for strategic reasons,  the fusion between medium-size  clusters may be more tactical depending on how a fight is evolving, and the fusion of small clusters may be ad hoc -- or any variant of these.

The starting equation is the rate of change of the number of clusters $n_s$ that contain $s$ fighters (size $s=1,2 \dots N$). This equals the number of new clusters of  size $s$ being created minus the number of existing clusters of size $s$ being lost. 
A cluster of size $s=s_1+s_2$ is created by fusion of two smaller clusters $s_1$ and $s_2$ (e.g. $6+5=11$ Fig. 1C). A cluster of size $s$ is lost either by its fusion with another cluster or its fission (e.g.  $11=1+1+\dots+1+1$ Fig. 1C). The algebra only requires the post-fission cluster fragments are small ( $<s_{\rm min}$), i.e.  it does not have to be total fission. 
If fighters' links/interactions are independent of the distance between them (e.g. unlimited online/phone use) a new link can emerge between any of the $s_1$ fighters in cluster 1 and any of the $s_2$ fighters in cluster 2, which fuses the two clusters to create a new cluster of size $s_1+s_2$. Hence the fusion rate depends on the product $s_1 s_2$. In the opposite case of  links/interactions being limited to fighters who are near each other (e.g. clusters on a two-dimensional checkerboard with no online/phone use) the fusion rate dependence becomes $s_1^{1/2} s_2^{1/2}$ because it only involves fighters on the cluster perimeters (clusters of size $s_1$ and $s_2$ have perimeters $s_1^{1/2}$ and $s_2^{1/2}$). Both cases have a pre-factor $F$: If the fighter force comprises only one adversarial species (e.g. Hamas fighters), $F$ is a number that depends on those Hamas fighters' average heterogeneity; but with $D>1$ adversarial species (e.g. Hamas, PIJ etc. fighters) $F$ becomes a $D$-dimensional matrix. 
Following reaction kinetics, the number of fighters $s$ in a cluster is likely to determine the total number of casualties $x$ in an event that involves that cluster: $x$ can include fighters on either side and civilians, and it can be any constant multiple or fraction of $s$. Taking the rate at which  clusters are involved in events as a constant (e.g.  because every fighter needs a similar time to recover or re-equip, regardless of its cluster size) the distribution $n_s$ will have the same mathematical form as the casualty distribution $n_x$, i.e.  the number of events with $x$ casualties. 

This mathematics yields three key predictions for any overall fight in which a fighter force is undergoing fusion and occasional total (or near total) fission in its fight against some typically strong opponent. Each of these can be seen and explored visually using the plug-and-play simulation: 

\vskip0.2in

\noindent {\bf Prediction 1: The number of events $n_x$ with $x$ casualties will have an approximate power-law distribution $x^{-\alpha}$ where $2.0\leq\alpha\leq 2.5$.} 
As proved mathematically in SI Eq. 37, $\alpha=2.5$ when fighter links/interactions are distance independent (e.g. unlimited online/phone use), and $\alpha=5/2-1/2\equiv 2.0$ when fighters only form links/interact with other fighters nearby (e.g. no online/phone use). Intermediate cases will lie between these values. This Prediction 1 is exact for $x\geq x_{\rm min}$ when the fighter force is large ($N\gg 1$). The plug-and-play simulation shows explicitly the $2.5$ case. Prediction 1 has a remarkable robustness to mathematical variations \cite{PRL2023,ben,scirep2013,Nature2009} which means it should also be a reasonable approximation more generally. 

\vskip0.2in

\noindent {\bf Prediction 2: If fission becomes extremely infrequent, a single giant cluster of fighters will suddenly emerge at a time $t_c$ which means high risk of a giant attack. Kinks in its growth curve mean that multiple adversarial species are involved and undergoing joint fusion-fission} (e.g. Hamas with PIJ and Fatah). The term `giant' is used in Physics to mean that the largest cluster's size has become a significant fraction ($G(t)$) of the entire fighter population $N$ \cite{PRL2023}. Hence it suddenly appears as a  surprise to any observer who does not know about the mesoscale cluster dynamics. Its mathematical `shock' shape is due to smaller but substantial clusters fusing together in quick succession  \cite{PRL2023,math,Ziff,ben}. 
The SI Secs. 3 and 4 derive exact formulae for its onset time $t_c$ and growth curve, but crudely $t_c\approx N/2F$. 

\vskip0.2in

\noindent {\bf Prediction 3: If multiple adversarial species (e.g.  Hamas, PIJ and Fatah) undergo joint fusion-fission with strong couplings (i.e. links/interactions) between them, their separate giant clusters will pile  together to create a single super-shock cluster -- and hence attack -- that will be stronger (more casualties) and arrive earlier than an attack like October 7.} 

\vskip1.0in

\subsection{Prediction 1 matches existing casualty data (Fig. 1C) and provides risk estimates for future casualties}

\vskip0.1in
In agreement with Prediction 1, Fig. 1C shows that all the casualty distributions from the Israel-Palestine region GED conflict and GTD terrorism data  \cite{Stijn} (see Methods) follow an approximate power-law distribution with $\alpha$ values broadly across the predicted range $2.0\leq \alpha\leq 2.5$. This result, and  Prediction 1 itself, involve no cherry-picking or fine-tuning of parameters, e.g. $N$, $F$ and the fusion/fission rates can have any values as long as $N\gg 1$ and fission is less frequent than fusion. The plug-and-play simulation shows the $2.5$ limit emerging explicitly. 

Approximate power-laws are known to arise for conflicts/terrorism and mechanisms have been proposed \cite{Richardson,scirep2013,Nature2009,Stijn,Oxford,Clauset,Dylan1,Dylan2,Oxford}. However, the comprehensive study in Ref. \cite{Stijn} showed that the empirical $\alpha$ values for conflicts/terrorism around the globe range from $1.37$ to $5.21$, which is well outside Prediction 1's narrow range $2.0\leq \alpha \leq 2.5$.  Richardson's original result of $\alpha=1.7$ across all wars, also falls outside  \cite{Richardson,Dylan1,Dylan2}. Furthermore, no other theoretical model has predicted this same narrow range $2.0\leq \alpha \leq 2.5$. This all suggests that Israel-Palestine violence represents a special subset of global violence, and that it is an archetypal example of fighter fusion-fission. 

Prediction 1 also predicts and explains a hidden post-October 7 shift in the specific Israel-Hamas etc. data-point from  $\alpha=2.6$ to  $\alpha=2.0$ (Fig. 1C). Given that the post-October 7 violence became focused in Gaza and hence a grid-like battlefield, and that long-range communications \cite{phone} became risky for Hamas etc. as well as being curtailed by Israel, Prediction 1 predicts a shift to $\alpha=2.0$ which is exactly as observed in the empirical casualty data (Fig. 1C).

Furthermore, Prediction 1 explains why the datapoint for pro-IS fighters online sits near $\alpha=2.5$ (blue box, Fig. 1C). Their online interactions are not restricted by geographical separation, hence Prediction 1  predicts $\alpha=2.5$ as observed empirically. The fact that the IS online and offline $\alpha$ values are so similar, suggests strong online-offline operational and organizational interplay.

Prediction 1 also allows calculation of concrete  risk estimates for future casualties in the Israel-Palestine region. For example, suppose the probability that a future event will produce $x_0$  casualties has been assessed as $p$. Prediction 1 shows the probability that it will instead produce $f$ times more casualties is $f^{-\alpha} p$. So if the chance of $x_0=100$ casualties is $10\%$, the chance of $1000$ casualties is 
$10^{-\alpha}.10=0.1\%$ for the case of nearby-fighter interactions  as currently in Gaza (i.e. $\alpha\rightarrow 2.0$). This is far higher than the value using the standard Gaussian risk assumption which would hence dangerously underestimate this risk.

 \vskip0.2in

\begin{figure}[ht]
    \centering
\includegraphics[width=0.45\linewidth]{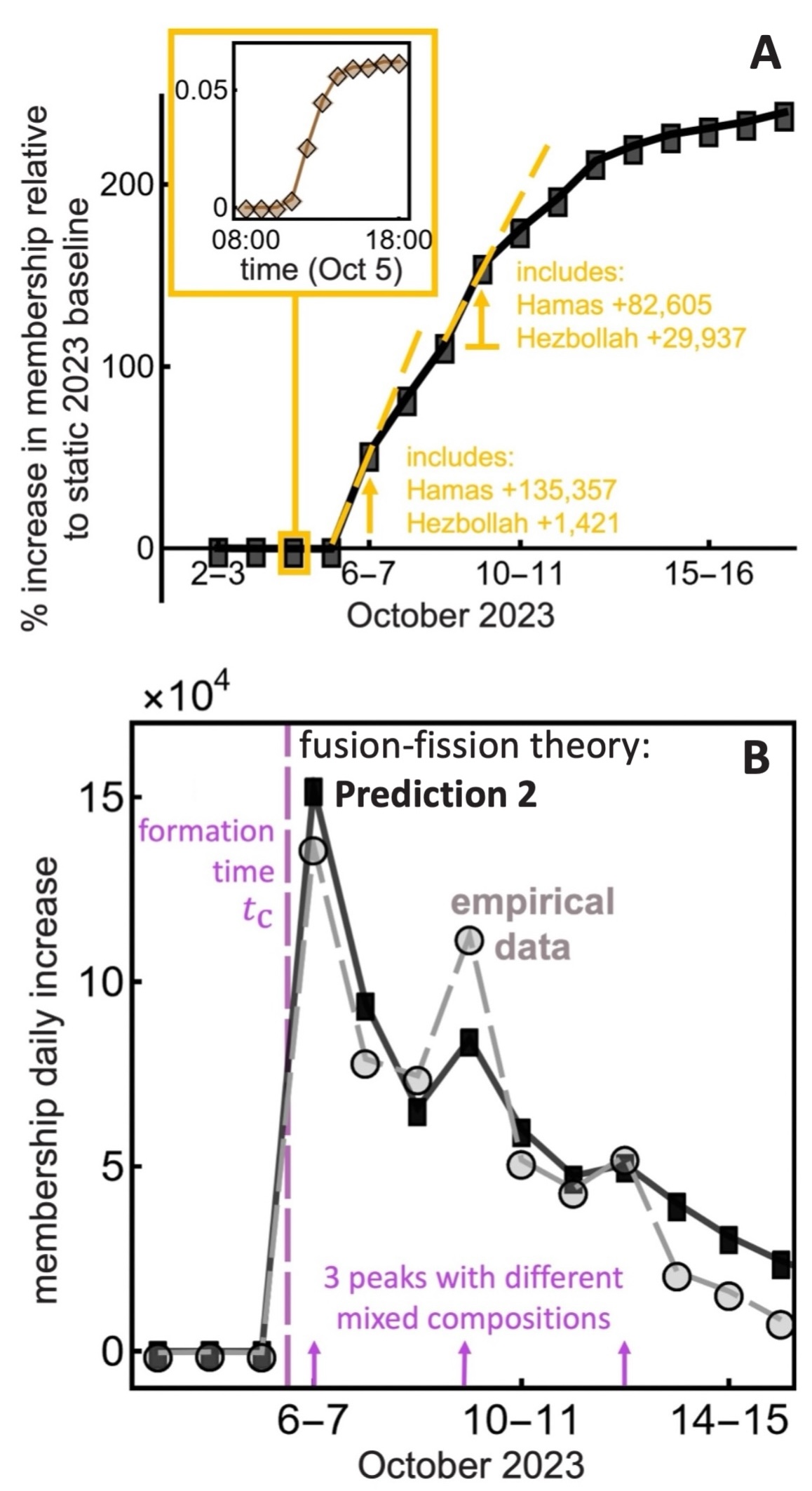}
    \caption{\bf Prediction 2 vs. empirical data. A: Empirical data for membership of anti-Israel military wing communities on Telegram. It plots Hamas and Hezbollah communities combined since they dominate, but these communities also include those with allegiance to PIJ etc. that may lack their own Telegram communities. Inset: glimpse of underlying mesoscale fusion. B: Rate of change of empirical data in A (dashed gray). Solid line is mathematical prediction (Prediction 2) for $D=3$ adversarial species (formulae in SI Secs. 3,4).}
    \label{fig:2}
\end{figure}

\subsection{Prediction 2 matches October 7 fighter data (Fig. 2)}

Prediction 2 matches fighters' collective behavior online around October 7: specifically, the massive growth in membership of anti-Israel military wing communities on Telegram (Fig. 2) which attract a diverse set of fighters akin to SI Fig. 1, i.e. they are not casual online users. The mathematical fit suggests $t_c$ is October 6 (see SI Sec. 1.5), i.e. a giant cluster of fighters surfaced the day before the attack, leaving the subset who were physically nearby several hours for the logistics of advancing into Israel. Figure 2A's inset shows evidence of smaller-scale fighter fusion prior to this giant cluster emergence, which is consistent with the cluster fusion-fission mathematics and can also be seen explicitly in the plug-and-play simulation. 

Prediction 2 also reproduces the details of the growth kinks, as shown in Fig. 2B which plots Fig. 2A's rate of change. It implies that the number of adversarial species that participated in the giant cluster and hence October 7 attack was at least three, i.e. $D\geq 3$. 

A simple -- but misleading -- takeaway would be that the October 7 attack happened because Israel  stopped generating effective fission events. However, the good fit for $D\geq 3$ adversarial species in Fig. 2 suggests the answer is more complex:  fighters from a minimum of three adversaries (e.g. Hamas, PIJ and others) were  undergoing fusion together in the same way at the same time. This explains why a multi-adversary attack force could so easily assemble, i.e. clusters could easily slot into each others' activity. It also explains why this multi-adversary assembly would have been missed by single-adversary  surveillance. Eye-witness accounts in SI Sec. 1.4 provide  additional independent support of this takeaway of  multi-adversary fusion of fighters.

\begin{figure}[ht]
    \centering
\includegraphics[width=0.45\linewidth]{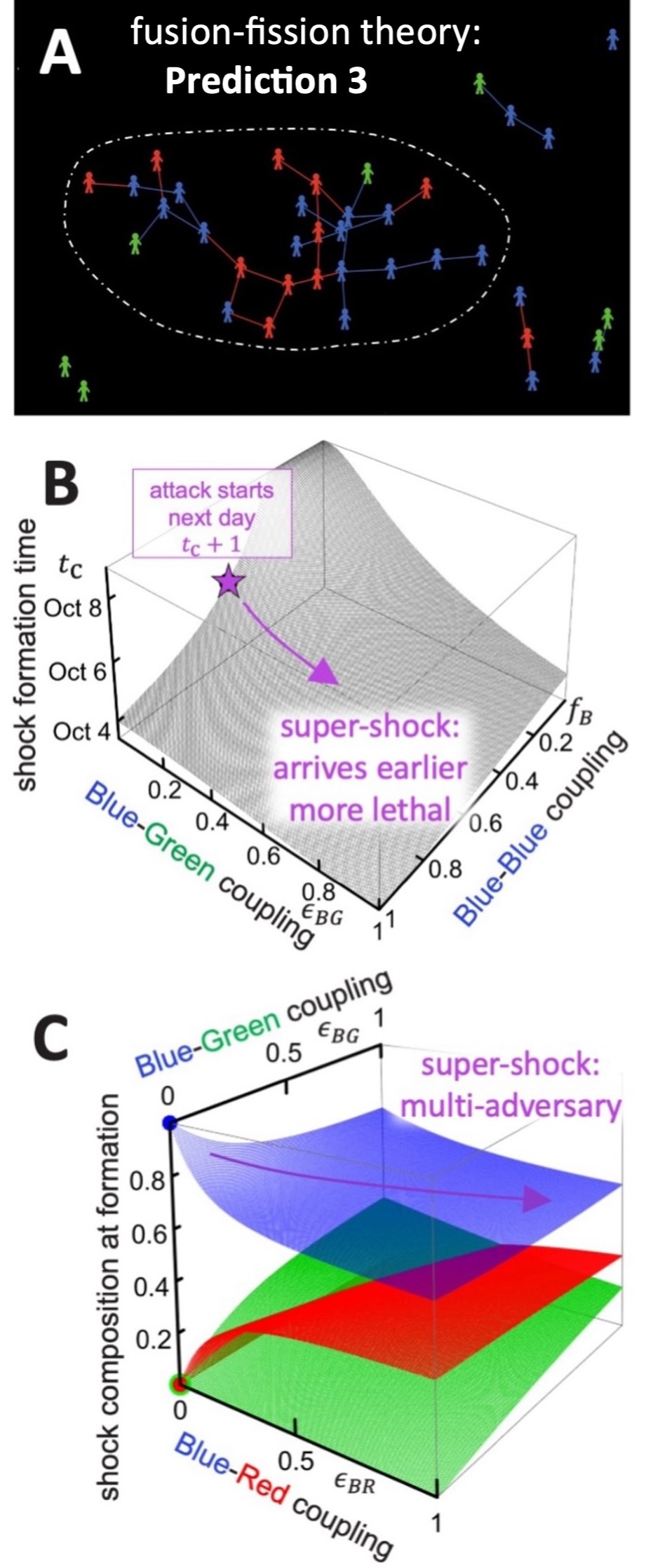}
    \caption{\bf Prediction 3 of a future super-shock attack. A: Snapshot from the plug-and-play simulation for $D=3$ adversarial species (e.g. blue Hamas, red PIJ, green Fatah). 
    The mathematical curves in B and C show the super-shock emerging as the couplings (links/interactions) between species become large. B shows its formation time and C shows its relative composition in terms of the $D=3$ adversarial species (blue, red, green as used in the plug-and-play simulation). The October 7 values (dots) are all non-zero but appear near 1 or 0 because of the large scales on the axes. These curves also show the October 7 attack would have occurred earlier and been more lethal if the couplings had been larger. Interventions can be explored using the plug-and-play simulation.}
    \label{fig:3}
\end{figure}

\vskip0.4in

\subsection{Prediction 3: future super-shock clusters and attacks (Fig. 3)}

Figure 3 shows how Prediction 3's super-shock cluster, and hence likely attack, emerge at large values of the couplings between adversarial species (i.e. many links/interactions). It also shows that the October 7 attack would have been progressively more lethal  and occurred earlier as these couplings increase. The plug-and-play simulation shows visually how the super-shock arises: giant clusters from each of the adversarial species pile up together. A key takeaway for policymakers is that any future such super-shock attack will not be attributable to a single adversarial species (e.g. Hamas). 

The super-shock formation time and hence earliest start of a super-shock attack approximates to:
 \begin{equation} \label{eq:effective_main}
    t_{\rm c}^{\rm super-shock} = \bigg( \frac{f}{f+(D-1)\epsilon}\bigg) t_{\rm c}^{\rm Oct\  7}
\end{equation}
given equal couplings between adversarial species  ($\epsilon>0$) and equal couplings within species ($f>0$). $t_{\rm c}^{\rm Oct\  7}$ is the formation time for an October 7-like attack. SI Secs. 3,4 derive  exact formulae. Equation 1 guarantees $t_c^{\rm super-shock} < t_c^{\rm Oct\ 7}$ and hence the super-shock attack  will arrive increasingly early as the  number of adversarial species $D$ increases or the  interactions between them $\epsilon$ increase.  
The key takeaway for policymakers is that the likelihood of a super-shock attack is {\em decreased} by  interventions that decrease the links/interactions between adversarial species $\epsilon$, and that this decrease can be estimated from Eq. 1 and explored explicitly using the plug-and-play simulation. 

The SI Sec. 5 proves mathematically that globalized (i.e.  multi-adversary) surveillance and intervention is indeed the best mitigation strategy for controlling fighter buildup. This can be verified visually using the plug-and-play simulation. 
\vskip0.1in

\section{Discussion}
Our findings present a unified quantitative description and explanation of Israel-Palestine region violence. 
They hence complement the rich body of existing work in conflict studies which addresses the politics, religion, history, economics, ideology and human psychology that undoubtedly play a key role in other aspects.  
Our `crude look at the whole' \cite{Gellmann,complex} allows us to calculate the consequences in a rigorous way, with the good empirical agreement (Figs. 1C,2B) suggesting that the net effect of the many missing details cancels out to some degree. But the question of why will require further study. 
Our findings also establish a concrete connection to fusion-fission studies across the animal kingdom \cite{Aureli} which suggests that the task of finding a lasting solution to human violence could benefit directly from combining their insights with those of conflict studies experts.

\vskip1.0in
\section*{Methods}
We provide an online {\bf plug-and-play simulation} of the paper's fusion-fission mathematics and its consequences including the Predictions 1,2 and 3, at \url{https://gwdonlab.github.io/netlogo-simulator/} which can be accessed and used in full by anyone, anytime using any browser -- including on a smartphone. It requires no coding or mathematical knowledge. It shows the fusion-fission process of single or multiple adversarial species, and it allows visual exploration of how the mathematical results and predictions arise and what they mean. It can also be used to explore interventions, run what-if-scenarios, and investigate their consequences. It is self-explanatory but SI Sec. 1.7 provides a brief starter manual in case useful.

The fighter data in Fig. 1A are from the state-of-the-art study led by John Horgan and Paul Gill (which lists N.F.J. as co-contributor) and are used with their kind permission
\cite{B2B}. The fighter data in Fig. 1B are from the state-of-the-art study of online fighter behavior led by N.F.J. and first reported in Ref. \cite{Science2016}. These references and SI Sec. 1 contain detailed discussions of how these data were collected etc. and some further examples of the fighters themselves. 

The conflict and terrorism data in Fig. 1C are from the Georeferenced Event Dataset (GED  \url{https://ucdp.uu.se/downloads/}) and the Global Terrorism Database (GTD  \url{https://www.start.umd.edu/gtd/}) respectively, as analyzed and discussed in depth in Ref. \cite{Stijn} which was co-authored by N.F.J. Ref. \cite{Stijn} also provides full replication code and statistical testing. It gives the error bars (standard deviation) in the $\alpha$ values and shows that these are typically small (hence we do not give them and instead refer to Ref. \cite{Stijn} for these values); and it gives the statistical goodness-of-fit values for the power-laws
and shows these are typically significant/large (hence again we refer to Ref. \cite{Stijn} for these values).
The dataset label `Israel: Palestine' from Ref. \cite{Stijn} was too vague and hence has been specified more precisely in Fig. 1C as `Israel: Hamas, PIJ, Fatah etc.'. Goodness-of-fit values for all the empirical power-laws are generally high (see Ref. \cite{Stijn} for details). The goodness-of-fit  value goes down for the post-October 2023  data-point featured in Fig. 1C but this is understandable given the preliminary form of the casualty data for the current war: see SI Sec. 1.6 for evidence of the crude estimation scheme being used for current Gaza casualties, specifically the tendency to report to the nearest factor of $10$. Our findings are unchanged if the casualty data are a systematic underestimate or overestimate by some factor, since taking the logarithm of a power-law distribution just means that factor adds to the intercept -- and $\alpha$ does not essentially depend on the intercept. 

Figure 2 data comes from militant communities on Telegram whose data is publicly available (see \url{https://ir.tgstat.com}). It shows the increase in members. The small, steady member attrition (e.g. moving to the Telegram equivalent of a private WhatsApp group, or un-joining) is not included because it is so small, steady, and has not changed significantly for years. 

The underlying mathematics is -- given its fusion-fission starting point (Sec. IIB) -- algebra that is agnostic of topic and sociopolicial labels, and is 100$\%$ reproducible using undergraduate skills without further debate. Hence we place it in the SI to avoid disrupting the flow of the main paper. It is written out in substantial detail in the SI so that anyone interested can see the rigorous foundations and then be taken through step-by-step.

\section*{Data Availability}
Data are publicly available as discussed in Methods, e.g. from Ref. \cite{Stijn}, GED and GTD websites, and  Telegram. The personal identities of the PIRA fighters in Fig. 1A, the pro-IS fighters in Fig. 1B, and the anti-Israel fighters in Fig. 2, are not available but nor are they needed or used since our study's methodology and  results only deal with the clusters that they form.

\section*{Code Availability}
The plug-and-play software for readers to scrutinize our fighter fusion-fission findings and explore interventions, is given online to use directly using any browser (\url{https://gwdonlab.github.io/netlogo-simulator/}). The code itself is at the same link (see tabs at bottom of webpage) together with a full explanation. Generic instructions for the underlying NetLogo machinery are publicly available through the NetLogo website (\url{https://ccl.northwestern.edu/netlogo/}) but the plug-and-play format itself is self-explanatory. SI Sec. 1.7 provides a brief starter manual. We are very grateful to Akshay Verma for helping set up this simulation. 
No special code was used to generate the results or the figures. The power-law results and code are given in Ref. \cite{Stijn} and on various other publicly available websites. Any type of open-source plotting software can be used that plots the equation solutions given in the SI.

\section*{Acknowledgements} N.F.J. is supported by U.S. Air Force Office of Scientific Research awards FA9550-20-1-0382 and FA9550-20-1-0383 and The Templeton Foundation. 

\section*{Author contributions}

F.Y.H, G.W. and N.F.J. conceived the empirical and theoretical connections to form the paper.  F.Y.H., P.M., D.J.R. and N.F.J. conducted the analysis. All authors analyzed the results.  All authors reviewed the manuscript. 

\section*{Competing interests}
The authors have no competing financial and/or non-financial interests in relation to the work described.

\section*{Supplementary Information (SI)}
Supplementary Information (SI) is available for this paper.
Correspondence and requests for materials should be addressed to N.F.J. (neiljohnson@gwu.edu)

\end{document}